# Mechanism-Dependent Descriptors for Oxygen Capacity in Perovskite Oxides

Shuiping Gong[1,2], Yi Li[1,2], Eric Isbrandt[3], Xiaoman Guo[3], Jie Su[1,2], Zhenhao Zhou[1,2], Xiaobo Liao[1,2], Jian Deng[1,2], Jiangang He[4,5], Tao Deng[1], Runxia Cai[6], Chaochao Dun[7], Zhenpeng Yao[1,2]*

[1]Center of Hydrogen Science and School of Materials Science and Engineering, Shanghai Jiao Tong University, Shanghai 200240, China

[2]Innovation Center for Future Materials, Zhangjiang Institute for Advanced Study, Shanghai Jiao Tong University, Shanghai 201203, China

[3]Acceleration Consortium, 700 University Avenue, Toronto, Ontario M7A 2S4, Canada

[4]Key Laboratory of Advanced Materials and Devices for Post-Moore Chips, Ministry of Education, University of Science and Technology Beijing, Beijing 100083, China

[5]School of Mathematics and Physics, University of Science and Technology Beijing, Beijing 100083, China

[6]School of Mechanical Engineering, Shanghai Jiao Tong University, Shanghai, China

[7]Department of Chemical and Biological Engineering, University at Buffalo, The State University of New York, Buffalo, NY 14260, United States of America

*E-mail: yaozhenpeng@gmail.com.

## Abstract

Perovskite oxides can reversibly accommodate substantial changes in oxygen stoichiometry, making them attractive for clean-energy technologies including chemical looping and oxygen storage. However, predicting oxygen capacity across diverse compositions remains challenging. Here, we combine experiments and first-principles calculations to establish composition-capacity relationships using the model perovskite series $Ln_xSr_{1-x}CoO_3$ ($x$ = 0.2~1.0, Ln = La–Lu). Increasing redox-inactive $Sr^{2+}$ content promotes high-valence $Co^{4+}$ formation, expanding the cationic redox reservoir and enhancing oxygen capacity. While oxygen-vacancy formation energy tracks this trend for light rare earths, net oxygen release counterintuitively decreases from La to Lu despite lower vacancy formation energies. We reveal that this behavior originates from an alternative charge-compensation pathway where lattice oxygen is partially oxidized to $O^{(2-\alpha)-}$-like species ($0<\alpha<2$). Heavy rare earths (Tb-Lu) preferentially stabilize these oxygen holes through distinct local bonding, with charge compensation involving both oxidized lattice oxygen and reduced cations, ultimately suppressing oxygen release. We identify the average metal-oxygen bond strength, quantified by the integrated crystal orbital Hamilton population (ICOHP), as a physically meaningful descriptor when anionic redox dominates. These findings reveal how distinct charge-compensation mechanisms determine the applicability of oxygen-capacity descriptors, establishing a mechanism-dependent framework for predicting and optimizing oxygen exchange in perovskite oxides.

## Introduction

Oxides capable of reversibly accommodating and releasing oxygen are central to a wide range of energy and environmental technologies, including chemical looping [1–3], solar thermochemical hydrogen production [4–6], energy storage [7–10], and gas separation [11–14]. In these applications, the amount of lattice oxygen that can be reversibly exchanged with the surrounding environment, *i.e.*, the oxygen capacity, governs key performance metrics such as oxygen yield [15,16], energy density [17,18], reaction efficiency [19,20], and long-term cycling stability [21,22]. Developing oxides with high oxygen capacity has therefore become an important objective in energy materials research. A variety of redox-active oxides have been explored for oxygen storage and release, including iron [23,24], manganese [25,26], cobalt [27,28], copper [29,30], and cerium oxides [31–33], as well as their multicomponent and high-entropy derivatives [34,35]. Among these materials, perovskite oxides ($ABO_3$) are particularly attractive because they combine exceptional compositional flexibility [36,37] with structural robustness [21,38], defect tolerance [39,40], and rapid oxygen transport kinetics [41]. The ability to independently tailor both A- [11,42] and B-site [21,40] chemistries enables tuning of oxygen nonstoichiometry and redox behavior [43,44], making perovskites one of the most versatile platforms for engineering oxygen capacity.

To improve its oxygen capacity, an oxide must simultaneously facilitate reversible lattice-oxygen exchange, provide sufficient charge-compensation capability, and enable rapid oxygen transport [15,45,46]. Previous studies have achieved these objectives by tuning the thermodynamics of oxygen accommodation and release [2,40,47], expanding the accessible cationic redox reservoir [48,49], and enhancing oxygen-ion diffusion [43,50], leading to the development of high-capacity oxides [16,51,52]. However, these strategies offer limited guidance for predicting oxygen capacity directly from composition. This contrasts with more established oxide reactions, such as the oxygen evolution reaction (OER) and oxygen reduction reaction (ORR), where electronic-structure descriptors have been used in general to guide catalyst design (*e.g.*, transition-metal $e_g$-filling and covalency) [53]. For oxygen capacity, oxygen vacancy formation energy has been the most widely adopted descriptor [26,54,55], yet its predictive capability is limited because oxygen capacity cannot be described solely by isolated vacancy formation [26,56–58]. Instead, it is governed by the coupled effects of cationic and anionic redox, oxygen vacancy interactions, structural evolution, and oxygen transport under various operating conditions [4,59]. Establishing a physically grounded descriptor that captures these coupled effects is therefore essential for predicting oxygen capacity and enabling the rational design of high-performance oxides.

Here, we address this challenge using the model perovskite series $Ln_xSr_{1-x}CoO_3$ ($x$ = 0.2~1.0, Ln = La–Lu). By combining systematic experiments with first-principles calculations, we establish composition-property relationships linking compositional chemistry, electronic structure, charge-compensation mechanisms, and oxygen capacity within this model system. We show that oxygen capacity is associated with distinct cationic- and anionic-redox pathways that emerge under different compositional regimes. When oxygen release is dominated by cobalt cationic redox, oxygen-vacancy formation energy successfully captures the oxygen-capacity trend. In contrast, when anionic redox becomes significant, average metal-oxygen bond strength, quantified by the integrated crystal orbital Hamilton population (ICOHP), provides a physically grounded descriptor by capturing the stabilization of oxidized lattice oxygen. These findings reveal the mechanistic origins of oxygen capacity variations and establish a mechanism-dependent approach for selecting physically relevant descriptors to guide composition optimization in perovskite oxygen carriers, and provide a predictive framework for the rational design of perovskite oxides and related oxygen-exchange materials.

**Redox-inactive alkaline-earth substitution enhances oxygen capacity**

Perovskite oxides provide an ideal platform for investigating composition and oxygen capacity relationships because their A- and B-site chemistries can be independently tailored while retaining the same crystal framework. As illustrated in **Fig. 1a**, a typical perovskite oxide consists of alkaline-earth, rare-earth, or alkali-metal cations occupying the A sites within a framework of corner-sharing $BO_6$ octahedra, where the B-site cations are typically transition metals. To establish how composition governs oxygen capacity, we selected $Ln_xSr_{1-x}CoO_3$ as a model system, in which Ln/Sr substitution tunes the A-site chemistry while Co serves as an active redox center for reversible oxygen accommodation. We first investigated how redox-inactive A-site substitution influences oxygen capacity by varying the Ln/Sr ratio (Ln=La, $x$ = 0.2~1.0), and the details of sample synthesis, characterization, and testing procedures are provided in **Supplementary Notes 1 and 2**. X-ray diffraction (**Fig. 1b**) confirms the formation of the perovskite phase across the composition range and reveals a gradual structural transition from the cubic $Pm\bar{3}m$ phase of $La_{0.2}Sr_{0.8}CoO_3$ to the rhombohedral $R\bar{3}c$ phase of $LaCoO_3$ as the Sr content decreases.

Thermogravimetric analysis (TGA) and mass spectrometry (MS)-based oxygen-production measurements reveal a pronounced dependence of oxygen capacity on A-site composition. As the initial oxygen nonstoichiometry ($\delta_0$) was not independently determined, TGA mass changes were converted into changes in oxygen nonstoichiometry ($\Delta\delta$) using molar masses calculated from the nominal stoichiometric compositions. Thus, $\Delta\delta$ quantifies oxygen exchange relative to the initial state of each sample rather than its absolute oxygen deficiency. Both the oxygen nonstoichiometry changes measured by TGA (**Fig. 1c**) and the oxygen-production rates and cumulative yields measured by MS (**Fig. 1d,e**) decrease systematically with increasing La content. The gradual mass loss during reduction is consistent with progressive lattice-oxygen release, rather than an abrupt transformation between distinct phases. Upon switching the atmosphere from Ar to $Ar/O_2$, all samples exhibit a clear mass gain, indicating oxygen reincorporation into the oxide lattice and providing evidence of reversible oxygen exchange under the applied conditions. Notably, samples with higher La content recover to a mass close to their initial values more rapidly, suggesting faster apparent reoxidation under the applied conditions. Post-reduction X-ray diffraction further confirms retention of the perovskite framework following oxygen release (**Supplementary Note 3** and **Fig. S1**). Together, these observations establish that increasing La content progressively suppresses oxygen release while preserving the underlying perovskite structure, raising the question of how A-site chemistry regulates the accessible oxygen-redox capacity.

To elucidate the origin of this compositional dependence, we first examined the cobalt oxidation states and their evolution during oxygen exchange. Substituting $Sr^{2+}$ with $La^{3+}$ lowers the average cobalt oxidation state required for charge neutrality, thereby reducing the initial $Co^{4+}/Co^{3+}$ ratio. Consistent with this expectation, Co 2*p* X-ray photoelectron spectroscopy (XPS) reveals a systematic decrease in the $Co^{4+}$ fraction with increasing La content (**Fig. 1g** and **Supplementary Fig. S2**). [60,61] The accompanying evolution of the lattice-oxygen O 1*s* binding energy further indicates changes in the local electronic environment (**Fig. 1f**).[62,63] The dominant O 1*s* feature at approximately 531.5 eV is attributed primarily to surface species, including carbonates and hydroxyls, whereas lattice oxygen is assigned to the lower-binding-energy component. Importantly, Co K-edge X-ray absorption near-edge structure (XANES) and Co 2*p* XPS measurements before and after reduction reveal a substantial decrease in cobalt oxidation state following oxygen release (**Supplementary Fig. S6**). These observations establish cobalt reduction as the principal cationic charge-compensation pathway. As the La content increases, the progressive depletion of high-valence $Co^{4+}$ reduces the accessible cationic redox reservoir, thereby limiting the amount of lattice oxygen that can be released. Consistently, La-rich compositions exhibit higher reduction-onset temperatures, indicating that oxygen release becomes less favorable under the investigated conditions.

First-principles calculations provide further insight into the electronic-structure changes underlying this behavior. As the La content increases, the O 2*p* states shift away from the Fermi level, accompanied by diminished oxygen-hole character near the valence-band maximum (**Fig. 1h**). Analysis of Co 3*d*-O 2*p* hybridization and integrated crystal orbital Hamilton population (ICOHP) further reveals systematic changes in the metal-oxygen bonding environment (**Fig. 1i** and **Supplementary Fig. S3**), consistent with the composition-dependent electronic structures observed experimentally. Bader charge analysis indicates that La and Sr remain essentially redox-inactive, whereas cobalt accommodates the dominant cationic charge redistribution associated with oxygen removal (**Fig. 1k** and **Supplementary Fig. S5**). Together with the spectroscopic evidence, these calculations support a predominantly cobalt-centered charge-compensation mechanism across the $La_xSr_{1-x}CoO_3$ series.

Having established the dominant charge-compensation pathway, we next examined whether oxygen-vacancy formation energy ($E_{V,O}$) could capture the experimentally observed oxygen-capacity trend. We calculated $E_{V,O}$ for inequivalent lattice-oxygen sites within each composition and found that the lowest vacancy formation energy increases monotonically with increasing La content (**Fig. 1j**, **Supplementary Note 6** and **Fig. S4**). This trend is consistent with the progressively lower oxygen yields measured by TGA, indicating that oxygen release becomes energetically less favorable as the accessible cobalt redox reservoir decreases. Within this compositional series, where oxygen removal is predominantly compensated by cobalt reduction, oxygen-vacancy formation energy therefore provides an effective descriptor of the experimentally observed oxygen-capacity trend.

Among the compositions investigated, $La_{0.2}Sr_{0.8}CoO_3$ exhibits the highest oxygen capacity, releasing 0.179 mol $O_2$ per mole of oxide during reduction at 1100 ℃ in Ar. This corresponds to approximately 72% of the idealized theoretical oxygen yield calculated for the stoichiometric $La_{0.2}Sr_{0.8}CoO_3 \rightarrow La_{0.2}Sr_{0.8}CoO_{2.5}$ conversion, indicating its great potential as an oxygen carrier. This percentage is referenced to the nominal stoichiometric composition, as the initial oxygen nonstoichiometry ($\delta_0$) was not experimentally determined. Overall, these results demonstrate that A-site substitution regulates oxygen capacity by modifying the accessible cobalt redox reservoir and the energetics of oxygen removal. More importantly, they establish a compositional regime in which cobalt-centered charge compensation enables oxygen-vacancy formation energy to capture oxygen-capacity trends, providing a reference for evaluating the applicability of this descriptor across other perovskite chemistries.

**Rare-earth substitutions reveal an unexpected oxygen capacity trend**

Having established that redox-inactive Sr substitution enhances oxygen capacity by expanding the cationic redox reservoir, we next investigated whether oxygen capacity could be further tailored through isovalent rare-earth substitution. Unlike La/Sr substitution, replacing $La^{3+}$ with other rare-earth cations preserves the nominal charge balance while systematically modifying the ionic radius and local structural environment through lanthanide contraction. The $Ln_{0.2}Sr_{0.8}CoO_3$ (Ln = La–Lu) series therefore provides a model system for examining the influence of rare-earth identity on oxygen capacity. Fourteen compositions, excluding radioactive Pm, were synthesized by solid-state reaction. X-ray diffraction confirms that all compositions predominantly adopt the perovskite structure (**Fig. 2a**). These structural variations provide a continuous compositional platform for probing oxygen accommodation across the lanthanide series.

Thermogravimetric analysis and mass spectrometry-based oxygen-production measurements reveal a pronounced and unexpected dependence of oxygen capacity on rare-earth identity. The oxygen nonstoichiometry changes ($\Delta\delta$) measured by TGA (**Fig. 2b**) and the oxygen-production rates measured by MS (**Fig. 2c**) both indicate an overall decrease in oxygen release from La to Lu. Despite the nominally isovalent nature of rare-earth substitution, progressively heavier rare-earth compositions release substantially less lattice oxygen under the same reduction conditions. As discussed above, $\Delta\delta$ represents the oxygen-content change relative to the initial state of each sample, and possible differences in initial oxygen nonstoichiometry across the series cannot be excluded. During subsequent reoxidation under $Ar/O_2$, all compositions exhibit mass recovery, indicating oxygen reincorporation under the investigated conditions. Heavy rare-earth compositions generally approach their initial masses sooner, primarily reflecting their smaller oxygen losses during reduction rather than necessarily faster intrinsic oxygen-uptake kinetics. Post-reduction X-ray diffraction further confirms retention of the perovskite framework, although structural distortions are observed following oxygen release (**Fig. 2f**, **Supplementary Note 4** and **Fig. S7**). Additional temperature-dependent measurements show that higher reduction temperatures generally increase oxygen yield but can compromise cycling stability (**Supplementary Note 5** and **Figs. S8-S13**). Together, these observations establish a strong dependence of oxygen capacity on rare-earth identity, beyond that expected from nominal A-site charge balance alone.

To examine the electronic origins of this compositional dependence, we characterized the cobalt and oxygen environments using X-ray photoelectron spectroscopy (XPS). The Co 2*p* spectra remain broadly similar across the rare-earth series (**Fig. 2e**), suggesting relatively small variations in the surface cobalt oxidation states despite substantial differences in oxygen capacity. In contrast, the lattice-oxygen O 1*s* component exhibits an overall shift toward higher binding energies from La to Lu, with some composition-dependent deviations (**Fig. 2d**). This evolution suggests systematic changes in the local electronic environment of lattice oxygen associated with rare-earth substitution. The contrasting responses of cobalt and oxygen indicate that the pronounced oxygen-capacity trend cannot be readily rationalized by changes in the initial cobalt oxidation state alone, pointing to additional electronic or bonding factors that influence oxygen release.

These observations raise a fundamental question: what governs oxygen capacity when nominal A-site charge balance and cobalt oxidation states remain largely unchanged? In the preceding La/Sr substitution series, $E_{V,O}$ successfully captured the experimentally observed oxygen-capacity trend under predominantly cobalt-centered charge compensation. We therefore calculated $E_{V,O}$ across the $Ln_{0.2}Sr_{0.8}CoO_3$ series to determine whether this descriptor remains applicable when the rare-earth identity, rather than the nominal A-site charge, is systematically varied.

**Distinct cationic and anionic charge-compensation pathways govern oxygen capacity**

First-principles calculations reveal that the $E_{V,O}$ decreases systematically from La to Lu (**Fig. 3a**), indicating that the formation of an isolated oxygen vacancy becomes progressively more favorable across the rare-earth series. This trend contrasts sharply with the experimentally observed decrease in oxygen capacity, demonstrating that $E_{V,O}$ alone cannot account for the oxygen-release behavior of these compositions. To understand this discrepancy, we first examined the electronic structures of the pristine perovskites. All compositions ($Ln_{0.25}Sr_{0.75}CoO_3$, Ln=La−Lu, approximating the experimental stoichiometry owing to computational supercell constraints) exhibit broadly similar projected densities of states, characterized by substantial Co 3*d*-O 2*p* hybridization and O 2*p* contributions near the valence-band maximum (**Supplementary Fig. S14**). These similarities suggest that the electronic structures of the pristine materials alone do not readily explain the pronounced differences in oxygen capacity. We therefore investigated the electronic response to oxygen removal, using $La_{0.25}Sr_{0.75}CoO_3$ as an initial reference before examining the composition-dependent charge redistribution across the rare-earth series. Partial charge-density (PARCHG) analysis and projected densities of states reveal changes in the O 2*p* electronic states near the Fermi level following vacancy formation (**Fig. 3b,c**), suggesting that oxygen removal also modifies the electronic environment of the remaining lattice oxygen. The accompanying shift of the lattice-oxygen O 1*s* component toward higher binding energies after reduction (**Supplementary Fig. S15**) is consistent with changes in the local oxygen electronic environment. Together, these observations support the development of partial oxygen-hole character, represented as $O^{(2-\alpha)-}$ ($0<\alpha<2$), in the remaining lattice oxygen, as reported for other oxygen-redox-active oxides.[27,62,64,65]

To determine how charge compensation evolves across the rare-earth series, Bader charge analysis was performed for the oxygen-deficient compositions. In light rare-earth compositions, oxygen removal is accompanied predominantly by charge redistribution toward cobalt cations, consistent with cobalt reduction from initially mixed $Co^{4+}/Co^{3+}$ states toward lower oxidation states (**Fig. 3d**). In contrast, heavy rare-earth compositions exhibit more pronounced changes in the electronic environments of both rare-earth cations and lattice oxygen. The larger reduction-induced changes in lanthanide Bader charges (**Fig. 3e**), together with the negligible response of Sr cations (**Supplementary Fig. S16**), indicate an increasing contribution from rare-earth-associated electronic states to charge redistribution. To experimentally assess this predicted difference in rare-earth electronic response, we examined the Ln core-level XPS spectra and $L_3$-edge XANES spectra before and after reduction. XPS reveals no detectable changes in the formal oxidation states of either light or heavy rare-earth cations, as evidenced by the largely preserved core-level line shapes, spin-orbit splittings, and satellite structures (**Supplementary Fig. S17** and **Supplementary Note 7**). Importantly, however, the absence of detectable formal valence changes does not preclude more subtle electronic redistribution. Indeed, La and Yb $L_3$-edge XANES measurements reveal distinct reduction-induced responses, with the Yb-containing composition exhibiting substantially more pronounced changes in white-line intensity and spectral broadening than its La counterpart (**Fig. 3f and 3g, Supplementary Note 8**). These spectral changes, which reflect modifications in the unoccupied electronic states and local coordination environment, provide experimental support for the stronger rare-earth-associated electronic redistribution predicted by the Bader charge analysis, without implying a transition between integer oxidation states. Concurrently, the oxygen Bader charges become less negative, with a larger fraction of oxygen sites exhibiting partial oxidation in heavy rare-earth compositions (**Fig. 3h**). [64,66] . Combined with the oxygen-hole states identified in the electronic-structure calculations, these results indicate that heavy rare-earth substitution promotes a distinct charge-compensation response involving more pronounced electronic redistribution among Co-, Ln-, and O-derived states and greater partial oxidation of the remaining lattice oxygen.

These findings suggest that oxygen capacity depends not only on the energetics of initial vacancy formation but also on the oxygen-deficient lattice's electronic response. Whereas oxygen removal in light rare-earth compositions is accommodated predominantly through cobalt reduction, heavy rare-earth compositions exhibit more pronounced rare-earth-associated charge redistribution and oxygen-hole formation. The resulting changes in electronic structure and local bonding may contribute to the stabilization of the remaining lattice oxygen, providing a mechanistic explanation for the suppressed oxygen release in heavy rare-earth perovskites despite their lower calculated vacancy formation energies. Thus, the experimentally accessible oxygen capacity cannot be described by isolated oxygen-vacancy formation energy alone but must also account for the composition-dependent electronic and bonding responses accompanying progressive oxygen removal. These findings further suggest that A-site cations, often treated as redox-inactive in perovskite oxides, can contribute to electronic charge redistribution during oxygen release without undergoing detectable formal valence transitions. Accordingly, the potential electronic involvement of A-site cations, including substituted transition metals, should not be overlooked when interpreting charge-compensation mechanisms or designing oxygen-exchange materials.

**Metal-oxygen bonding provides a descriptor for oxygen capacity**

The preceding results reveal that oxygen-vacancy formation energy alone cannot explain the oxygen-capacity trend across the rare-earth series, highlighting the importance of composition-dependent charge compensation during oxygen removal. Because these electronic responses are closely coupled to the local metal-oxygen bonding environment, we investigated whether metal-oxygen bonding could provide a complementary descriptor for oxygen capacity, particularly when anionic redox becomes significant.

To quantify the evolution of metal-oxygen bonding, we performed crystal orbital Hamilton population (COHP) analysis, which resolves the energy-dependent bonding and antibonding interactions between neighboring atoms. Although the perovskite compositions exhibit broadly similar Co 3*d*-O 2*p* hybridization and antibonding contributions near the Fermi level (**Supplementary Fig. S18**), the bonding-related electronic states evolve systematically across the rare-earth series. We therefore calculated the integrated COHP (ICOHP), which quantifies occupied bonding interactions by integrating the energy-resolved COHP up to the Fermi level. For each lattice-oxygen site, we summed the ICOHP contributions from its neighboring Co, Ln, and Sr cations to characterize the local metal-oxygen bonding environment (**Fig. 4a**). The resulting average ICOHP becomes progressively more negative from La to Lu in both pristine and oxygen-deficient structures (**Fig. 4b**), indicating stronger overall metal-oxygen bonding interactions across the rare-earth series. Oxygen-vacancy formation also broadens the distribution of local ICOHP values, reflecting increased bonding heterogeneity following oxygen removal. Notably, the progressive strengthening of metal-oxygen bonding correlates with the experimentally observed decrease in oxygen capacity, despite the opposite trend in oxygen-vacancy formation energy. This correspondence suggests that the local bonding environment contributes to the stabilization of residual lattice oxygen, providing a complementary perspective on the composition-dependent oxygen-release behavior.

We further examined the local coordination environments of oxygen sites exhibiting pronounced oxygen-hole character following vacancy formation (**Fig. 4c**). In light rare-earth compositions, these sites are predominantly associated with $Co_2LnSr_3$ coordination environments. In contrast, heavy rare-earth compositions exhibit a broader range of local coordination motifs, including $Co_2Ln_2Sr_2$ and $Co_2Sr_4$ environments in addition to $Co_2LnSr_3$. This increased diversity of local coordination environments is consistent with the greater electronic and bonding heterogeneity identified in the oxygen-deficient heavy rare-earth perovskites. Together with the ICOHP and charge-density analyses, these observations suggest that composition-dependent local bonding environments contribute to the stabilization of partially oxidized lattice oxygen, thereby influencing the extent of oxygen release.

Taken together, these results identify average metal-oxygen bonding strength, quantified by ICOHP, as a physically grounded descriptor that captures the experimentally observed oxygen-capacity trend across the rare-earth series. [67,68] Unlike oxygen-vacancy formation energy, which describes the energetics of a specified oxygen-removal process, ICOHP characterizes the collective metal-oxygen bonding interactions that influence the electronic environment and stability of the remaining lattice oxygen. [69] Its correlation with oxygen capacity is particularly relevant in compositions exhibiting pronounced oxygen-hole character, where isolated vacancy formation energies do not reproduce the experimental trend. More broadly, the contrasting behavior of the La/Sr and rare-earth substitution series demonstrates that the predictive relevance of a descriptor depends on the underlying charge-compensation mechanism. This mechanism-dependent framework provides a basis for selecting physically meaningful descriptors and guiding the compositional design of redox-active perovskite oxides with tailored oxygen capacities.

## Conclusions

In summary, we establish a mechanism-dependent framework for understanding oxygen-capacity variations in perovskite oxides. Using $Ln_xSr_{1-x}CoO_3$ as a model system, we demonstrate that increasing $Sr^{2+}$ content enhances oxygen capacity by expanding the accessible $Co^{4+}$ redox reservoir, whereas rare-earth substitution produces a decreasing oxygen-capacity trend from La to Lu despite progressively lower oxygen-vacancy formation energies. Electronic-structure calculations, supported by spectroscopic characterization, reveal distinct charge-compensation responses involving Co-, Ln-, and O-derived states. Notably, the electronic response of rare-earth A-site cations, despite the absence of detectable formal valence transitions, challenges the assumption that these cations are electronically inactive during oxygen release. Oxygen-vacancy formation energy captures oxygen-capacity trends under predominantly cobalt-centered charge compensation, whereas average metal-oxygen bonding strength, quantified by integrated crystal orbital Hamilton population (ICOHP), correlates with oxygen capacity across the rare-earth series exhibiting more pronounced oxygen-hole character. These findings highlight the mechanism-dependent applicability of oxygen-capacity descriptors rather than a single universal metric. More broadly, our results provide a mechanistic basis for understanding composition-dependent oxygen capacity and guiding descriptor selection in the optimization of perovskite oxygen-exchange materials.

## Data Availability

The data supporting the findings of this study, including the experimental and computational data, are available within the article and its Supplementary Information. Additional data are available from the corresponding author upon reasonable request.

## Acknowledgements

Z.Y., S.G., Y.L., J.S., X.L., Z.Z., and J.D. were supported by the National Natural Science Foundation of China (grant number 52373228).

## Author contributions

Z.Y.: Conceptualization, Methodology, Funding acquisition, Project administration, Supervision. S.G.: Investigation, Data curation, Formal analysis, Visualization, Writing – original draft. Y.L.: Software. J.S.: Software. Z.Z.: Software. X.L.: Data curation. J.D.: Data curation. E. I.: Data curation. X.G.: Data curation. All authors: Writing – review & editing.

## Competing interests

The authors declare no competing interests.

## References

1. Jacob, M., Nguyen, H., Neurock, M. & Bhan, A. Selective chemical looping combustion of terminal alkynes in mixtures with alkenes. *J. Am. Chem. Soc.* **148**, 3139–3147 (2026).
2. Zeng, L., Cheng, Z., Fan, J. A., Fan, L.-S. & Gong, J. Metal oxide redox chemistry for chemical looping processes. *Nat. Rev. Chem.* **2**, 349–364 (2018).
3. Zhu, X., Imtiaz, Q., Donat, F., Müller, C. R. & Li, F. Chemical looping beyond combustion – a perspective. *Energy Environ. Sci.* **13**, 772–804 (2020).
4. Wexler, R. B. *et al.* Multiple and nonlocal cation redox in Ca-Ce-Ti-Mn oxide perovskites for solar thermochemical applications. *Energy Environ. Sci.* **16**, 2550–2560 (2023).
5. Perry, J. *et al.* Discovery of materials for solar thermochemical hydrogen combining machine learning, computational chemistry, experiments and system simulations. *NPJ Comput. Mater.* **11**, 247 (2025).
6. Feng, Z. *et al.* Recent advances in cerium oxide redox cycle for solar thermochemical hydrogen production. *Fuel* **407**, 137499 (2026).
7. Cai, R. *et al.* Accelerated Perovskite Oxide Development for Thermochemical Energy Storage by a High-Throughput Combinatorial Approach. *Adv. Energy Mater.* **13**, 2203833 (2023).
8. Carrillo, A. J., González-Aguilar, J., Romero, M. & Coronado, J. M. Solar Energy on Demand: A Review on High Temperature Thermochemical Heat Storage Systems and Materials. *Chem. Rev.* **119**, 4777–4816 (2019).
9. Xu, H. J. *et al.* Progress on thermal storage technologies with high heat density in renewables and low carbon applications: Latent and thermochemical energy storage. *Renew. Sustain. Energy Rev.* **215**, 115587 (2025).
10. Park, G. T. *et al.* Zero-strain Mn-rich layered cathode for sustainable and high-energy next-generation batteries. *Nat. Energy* **10**, 1215–1225 (2025).
11. Ogawa, S. *et al.* New Triclinic Perovskite-Type Oxide $Ba_5CaFe_4O_{12}$ for Low-Temperature Operated Chemical Looping Air Separation. *J. Am. Chem. Soc.* **145**, 22788–22795 (2023).
12. Sunarso, J., Hashim, S. S., Zhu, N. & Zhou, W. Perovskite oxides applications in high temperature oxygen separation, solid oxide fuel cell and membrane reactor: A review. *Prog. Energy Combust. Sci.* **61**, 57–77 (2017).
13. Zhou, X. *et al.* A ceria-protected three-layered oxygen transport membrane for stable and efficient membrane reactors. *Chem. Eng. J.* **515**, 163358 (2025).
14. Xiang, Z. *et al.* Oxygen permeability and stability in the entropy-stabilized Co-based perovskite oxygen permeable membranes. *J. Memb. Sci.* **739**, 124936 (2026).
15. Liu, C. *et al.* Perovskite Oxide Materials for Solar Thermochemical Hydrogen Production from Water Splitting through Chemical Looping. *ACS Catal.* **14**, 14974–15013 (2024).
16. Qian, X. *et al.* Outstanding Properties and Performance of $CaTi_{0.5}Mn_{0.5}O_{3-\delta}$ for Solar-Driven Thermochemical Hydrogen Production. *Matter* **4**, 688–708 (2021).
17. Huang, W. *et al.* A Quasi-Ordered Mn-Rich Cathode with Highly Reversible Oxygen Anion Redox Chemistry. *J. Am. Chem. Soc.* **147**, 26218–26225 (2025).
18. Wang, D., Wang, W., Ma, J. & Zhao, H. B-site Ni-doping and $MgO/TiO_2$ modified $CaMnO_{3-\delta}$ for thermal energy storage. *J. Mater. Chem. A* **14**, 29251–29265 (2026).
19. Abdulmoez, M., Othman, A., Linjawi, M. & Al-Qutub, A. Solar fuels via methane valorization: Thermochemical pathways, unified metrics, and techno-economic perspectives. *Energy Environ. Mater.* **9**, e70331 (2026).
20. Tran, J. T. *et al.* Pressure-enhanced performance of metal oxides for thermochemical water and carbon dioxide splitting. *Joule* **7**, 1759–1768 (2023).
21. Zhang, D. *et al.* Compositionally Complex Perovskite Oxides for Solar Thermochemical Water Splitting. *Chem. Mater.* **35**, 1901–1915 (2023).
22. Chung, C., Qin, L., Shah, V. & Fan, L. S. Chemically and physically robust, commercially-viable iron-based composite oxygen carriers sustainable over 3000 redox cycles at high temperatures for chemical looping applications. *Energy Environ. Sci.* **10**, 2318–2323 (2017).

23. Liu, Y. *et al.* Near 100% CO selectivity in nanoscaled iron-based oxygen carriers for chemical looping methane partial oxidation. *Nat. Commun.* **10**, 5503 (2019).
24. Zhai, S. *et al.* The use of poly-cation oxides to lower the temperature of two-step thermochemical water splitting. *Energy Environ. Sci.* **11**, 2172–2178 (2018).
25. Rydén, M., Leion, H., Mattisson, T. & Lyngfelt, A. Combined oxides as oxygen-carrier material for chemical-looping with oxygen uncoupling. *Appl. Energy* **113**, 1924–1932 (2014).
26. Mishra, A., Li, T., Li, F. & Santiso, E. E. Oxygen Vacancy Creation Energy in Mn-Containing Perovskites: An Effective Indicator for Chemical Looping with Oxygen Uncoupling. *Chem. Mater.* **31**, 689–698 (2019).
27. Mefford, J. T. *et al.* Water electrolysis on $La_{1-x}Sr_xCoO_{3-\delta}$ perovskite electrocatalysts. *Nat. Commun.* **7**, 11053 (2016).
28. Itoh, T. *et al.* Correlation between structure and mixed ionic-electronic conduction mechanism for $(La_{1-x}Sr_x)CoO_{3-\delta}$ using synchrotron X-ray analysis and first principles calculations. *J. Mater. Chem. A* **3**, 6943–6953 (2015).
29. Siriwardane, R. *et al.* 50-kWth methane/air chemical looping combustion tests with commercially prepared $CuO$-$Fe_2O_3$-alumina oxygen carrier with two different techniques. *Appl. Energy* **213**, 92–99 (2018).
30. Imtiaz, Q., Broda, M. & Müller, C. R. Structure–property relationship of co-precipitated Cu-rich, $Al_2O_3$- or $MgAl_2O_4$-stabilized oxygen carriers for chemical looping with oxygen uncoupling (CLOU). *Appl. Energy* **119**, 557–565 (2014).
31. Montini, T., Melchionna, M., Monai, M. & Fornasiero, P. Fundamentals and Catalytic Applications of $CeO_2$ -Based Materials. *Chem. Rev.* **116**, 5987–6041 (2016).
32. Haribal, V. P. *et al.* Modified Ceria for "Low-Temperature" $CO_2$ Utilization: A Chemical Looping Route to Exploit Industrial Waste Heat. *Adv. Energy Mater.* **9**, 1901963 (2019).
33. Scheffe, J. R. & Steinfeld, A. Oxygen exchange materials for solar thermochemical splitting of $H_2O$ and $CO_2$: a review. *Mater. Today* **17**, 341–348 (2014).
34. Chen, Z., Ma, B., Dang, C., Song, J. & Zhou, Y. High-Entropy Spinel Oxide Nanostructures as Stable Cathodes for Solid Oxide Fuel Cells. *Nano Lett.* **25**, 6421–6428 (2025).
35. Liu, C. *et al.* Manganese-based A-site high-entropy perovskite oxide for solar thermochemical hydrogen production. *J. Mater. Chem. A* **12**, 3910–3922 (2023).
36. Yang, K. & Li, F. Computationally accelerated discovery of mixed metal compounds for chemical looping combustion and beyond. *Energy Environ. Sci.* **18**, 10036–10047 (2025).
37. Vieten, J. *et al.* Materials design of perovskite solid solutions for thermochemical applications. *Energy Environ. Sci.* **12**, 1369–1384 (2019).
38. Zhu, X., Li, K., Neal, L. & Li, F. Perovskites as Geo-inspired Oxygen Storage Materials for Chemical Looping and Three-Way Catalysis: A Perspective. *ACS Catal.* **8**, 8213–8236 (2018).
39. Wexler, R. B., Gautam, G. S., Stechel, E. B. & Carter, E. A. Factors Governing Oxygen Vacancy Formation in Oxide Perovskites. *J. Am. Chem. Soc.* **143**, 13212–13227 (2021).
40. Wang, X. *et al.* High-throughput oxygen chemical potential engineering of perovskite oxides for chemical looping applications. *Energy Environ. Sci.* **15**, 1512–1528 (2022).
41. Barcellos, R. D., Sanders, M. D., Tong, J., McDaniel, A. H. & O'Hayre, R. P. $BaCe_{0.25}Mn_{0.75}O_{3-\delta}$ - promising perovskite-type oxide for solar thermochemical hydrogen production. *Energy Environ. Sci.* **11**, 3256–3265 (2018).
42. Ji, Q., Bi, L., Zhang, J., Cao, H. & Zhao, X. S. The role of oxygen vacancies of $ABO_3$ perovskite oxides in the oxygen reduction reaction. *Energy Environ. Sci.* **13**, 1408–1428 (2020).
43. He, J. *et al.* Subsurface A-site vacancy activates lattice oxygen in perovskite ferrites for methane anaerobic oxidation to syngas. *Nat. Commun.* **15**, 5422 (2024).
44. Mueller, D. N., Machala, M. L., Bluhm, H. & Chueh, W. C. Redox activity of surface oxygen anions in oxygen-deficient perovskite oxides during electrochemical reactions. *Nat. Commun.* **6**, 6097 (2015).
45. Long, T. *et al.* Modulating Activity of Lattice Oxygen of $ABO_3$ Perovskite Oxides in Redox Reactions: A Review. *ACS Appl. Mater. Interfaces* **17**, 20590–20612 (2025).

46. Bozal-Ginesta, C. *et al.* Performance Prediction of High-Entropy Perovskites $La_{0.8}Sr_{0.2}Mn_xCo_yFe_zO_3$ with Automated High-Throughput Characterization of Combinatorial Libraries and Machine Learning. *Adv. Mater.* **36**, 2407372 (2024).
47. Qian, X. *et al.* Favorable Redox Thermodynamics of $SrTi_{0.5}Mn_{0.5}O_{3-\delta}$ in Solar Thermochemical Water Splitting. *Chem. Mater.* **32**, 9335–9346 (2020).
48. Liu, L., Taylor, D. D., Rodriguez, E. E. & Zachariah, M. R. Influence of transition metal electronegativity on the oxygen storage capacity of perovskite oxides. *Chem. Commun.* **52**, 10369–10372 (2016).
49. Klimkowicz, A., Świerczek, K., Takasaki, A. & Dabrowski, B. Oxygen storage capability in Co- and Fe-containing perovskite-type oxides. *Solid State Ion.* **257**, 23–28 (2014).
50. Mayeshiba, T. T. & Morgan, D. D. Factors controlling oxygen migration barriers in perovskites. *Solid State Ion.* **296**, 71–77 (2016).
51. Capstick, S., Bulfin, B., Naik, J. M., Gigantino, M. & Steinfeld, A. Oxygen separation via chemical looping of the perovskite oxide $Sr_{0.8}Ca_{0.2}FeO_3$ in packed bed reactors for the production of nitrogen from air. *Chem. Eng. J.* **452**, 139289 (2023).
52. Cai, R. *et al.* High-throughput design of complex oxides as isothermal, redox-activated $CO_2$ sorbents for green hydrogen generation. *Energy Environ. Sci.* **17**, 6279–6290 (2024).
53. Suntivich, J. *et al.* Design principles for oxygen-reduction activity on perovskite oxide catalysts for fuel cells and metal-air batteries. *Nat. Chem.* **3**, 546–550 (2011).
54. Emery, A. A., Saal, J. E., Kirklin, S., Hegde, V. I. & Wolverton, C. High-Throughput Computational Screening of Perovskites for Thermochemical Water Splitting Applications. *Chem. Mater.* **28**, 5621–5634 (2016).
55. Deml, A. M. *et al.* Tunable Oxygen Vacancy Formation Energetics in the Complex Perovskite Oxide $Sr_xLa_{1-x}Mn_yAl_{1-y}O_3$. *Chem. Mater.* **26**, 6595–6602 (2014).
56. Park, J. *et al.* Accurate prediction of oxygen vacancy concentration with disordered A-site cations in high-entropy perovskite oxides. *npj Comput. Mater.* **9**, 29 (2023).
57. Millican, S. L. *et al.* Predicting Oxygen Off-Stoichiometry and Hydrogen Incorporation in Complex Perovskite Oxides. *Chem. Mater.* **34**, 510–518 (2022).
58. Mebane, D. S. & De Souza, R. A. A generalised space-charge theory for extended defects in oxygen-ion conducting electrolytes: from dilute to concentrated solid solutions. *Energy Environ. Sci.* **8**, 2935–2940 (2015).
59. Zhang, X. *et al.* $FeO_6$ Octahedral Distortion Activates Lattice Oxygen in Perovskite Ferrite for Methane Partial Oxidation Coupled with $CO_2$ Splitting. *J. Am. Chem. Soc.* **142**, 11540–11549 (2020).
60. Zhao, Y.-N. *et al.* An interface engineering strategy of $MoS_2$/perovskite oxide as a bifunctional catalyst to boost overall water splitting. *J. Mater. Chem. A* **12**, 8757–8768 (2024).
61. Pan, Y. *et al.* Direct evidence of boosted oxygen evolution over perovskite by enhanced lattice oxygen participation. *Nat. Commun.* **11**, 2002 (2020).
62. Shen, Z. *et al.* Correlating the electronic structure of perovskite $La_{1-x}Sr_xCoO_3$ with activity for the oxygen evolution reaction: The critical role of Co 3d hole state. *J. Energy Chem.* **65**, 637–645 (2022).
63. Cheng, X. *et al.* Oxygen Evolution Reaction on $La_{1-x}Sr_xCoO_3$ Perovskites: A Combined Experimental and Theoretical Study of Their Structural, Electronic, and Electrochemical Properties. *Chem. Mater.* **27**, 7662–7672 (2015).
64. Grimaud, A. *et al.* Activating lattice oxygen redox reactions in metal oxides to catalyse oxygen evolution. *Nat. Chem.* **9**, 457–465 (2017).
65. Assat, G. & Tarascon, J. M. Fundamental understanding and practical challenges of anionic redox activity in Li-ion batteries. *Nat. Energy* **3**, 373–386 (2018).
66. Seo, D. H. *et al.* The structural and chemical origin of the oxygen redox activity in layered and cation-disordered Li-excess cathode materials. *Nat. Chem.* **8**, 692–697 (2016).
67. Zhang, R. *et al.* Tailoring Catalytic and Oxygen Release Capability in $LaFe_{1-x}Ni_xO_3$ to Intensify Chemical Looping Reactions at Medium Temperatures. *ACS Catal.* **14**, 7771–7787 (2024).
68. Chen, S. *et al.* Modulating Lattice Oxygen in Dual-Functional Mo-V-O Mixed Oxides for Chemical Looping Oxidative Dehydrogenation. *J. Am. Chem. Soc.* **141**, 18653–18657 (2019).

69. Kuznetsov, D. A. *et al.* Tuning Redox Transitions via Inductive Effect in Metal Oxides and Complexes, and Implications in Oxygen Electrocatalysis. *Joule* **2**, 225–244 (2018).

# Figures

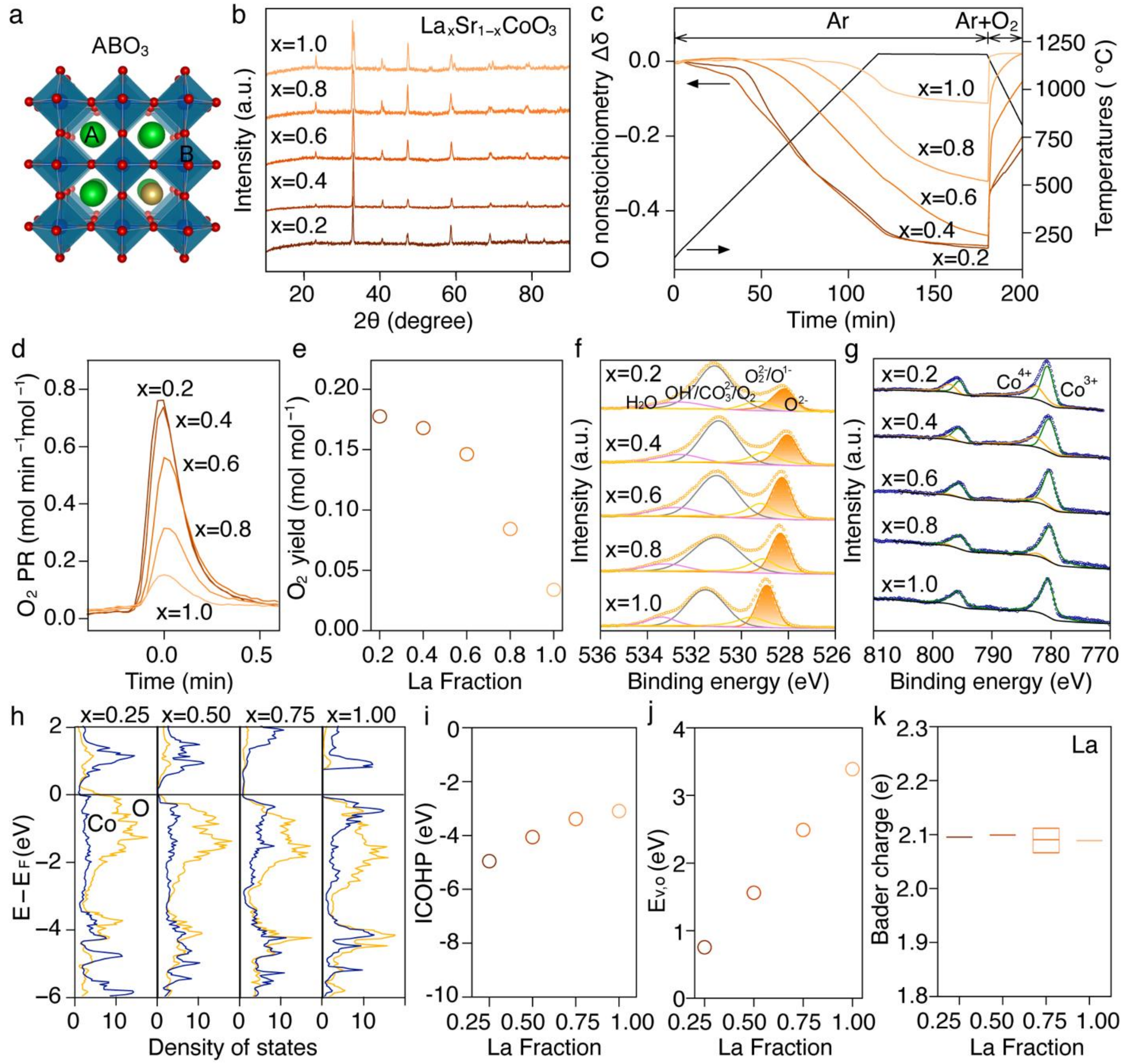

**Figure 1. Redox-inactive alkaline-earth substitution enhances oxygen capacity.** **a**, Schematics depicting the structure of $ABO_3$ perovskite oxides, A=alkaline-earth, rare-earth, or alkali metals. **b**, XRD patterns of pristine $La_xSr_{1-x}CoO_3$ perovskites. **c**, Oxygen nonstoichiometry change ($\Delta\delta$) measured by TGA. **d**, $O_2$ production rate of $La_xSr_{1-x}CoO_3$ perovskites reduced at 1100 ℃ and Ar flow. **e**, Cumulative $O_2$ yields. **f** and **g**, the O 1*s* and Co 2*p* core-level XPS spectra of $La_xSr_{1-x}CoO_3$ perovskites. **h**, The pDOS of Co 3*d* and O 2*p* of $La_xSr_{1-x}CoO_3$ perovskites. **i**, ICOHP of M-O bonding in $La_xSr_{1-x}CoO_3$ perovskites as a function of *x*. **j**, Oxygen vacancy formation energy ($E_{V,O}$) of $La_xSr_{1-x}CoO_3$ perovskites as a function of *x*. **k**, Bader charge of La cations in $La_xSr_{1-x}CoO_3$ perovskites.

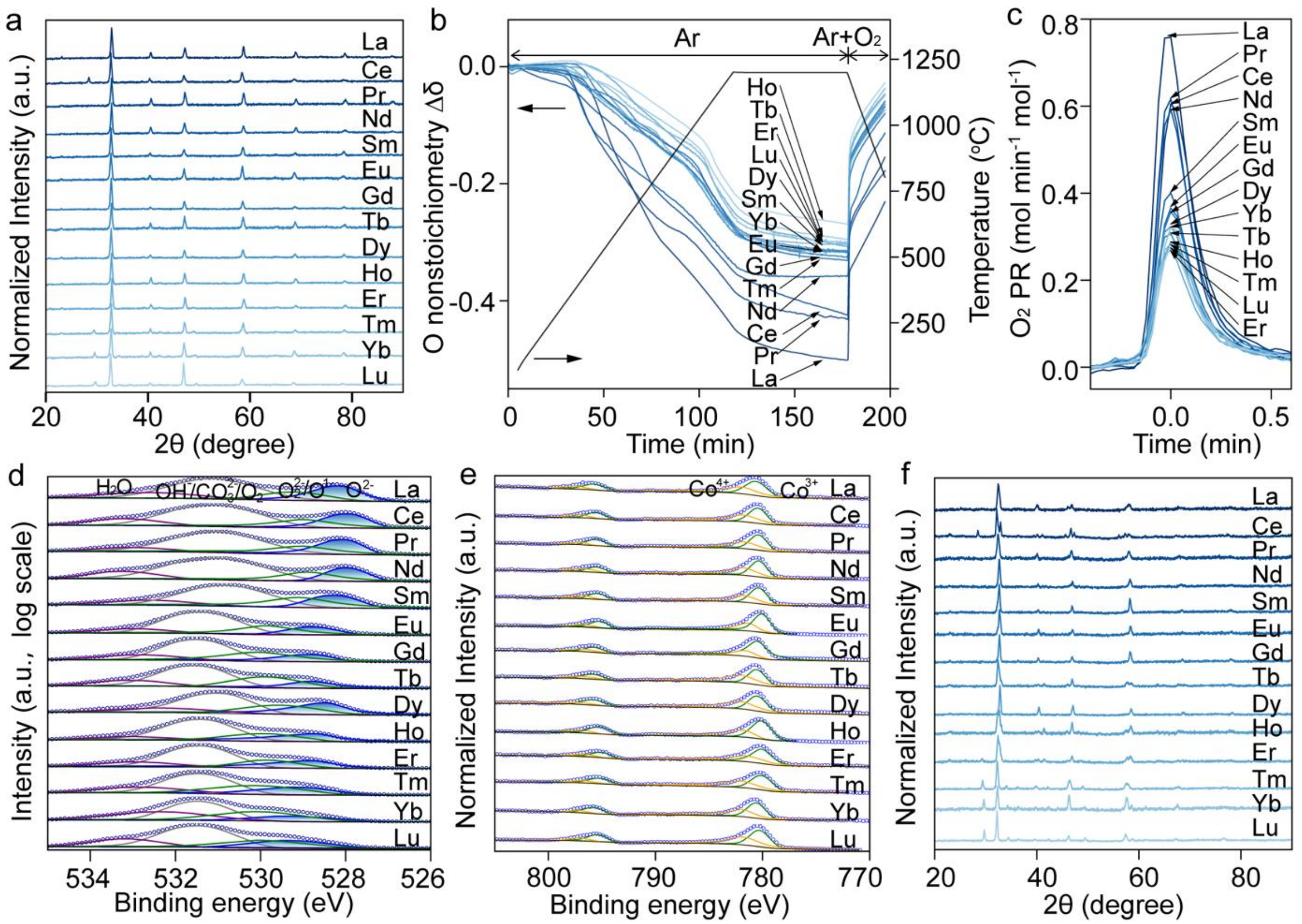

**Figure 2. Rare-earth substitution reveals the limitations of oxygen vacancy formation energy. a**, The XRD patterns of $Ln_{0.2}Sr_{0.8}CoO_{3-\delta}$ perovskites. **b**, Oxygen nonstoichiometry change (Δδ) measured by TGA. **c**, $O_2$ production rate of $Ln_{0.2}Sr_{0.8}CoO_{3-\delta}$ perovskites reduced at 1100 ℃ and Ar flow. **d** and **e**, The O 1*s* and Co *2p* core-level XPS spectra of $Ln_{0.2}Sr_{0.8}CoO_{3-\delta}$ perovskites. The O 1s spectra are displayed using a logarithmic intensity scale to visualize the weaker lattice-oxygen component better. **f**, The XRD patterns of the reduced $Ln_{0.2}Sr_{0.8}CoO_{3-\delta}$ perovskites.

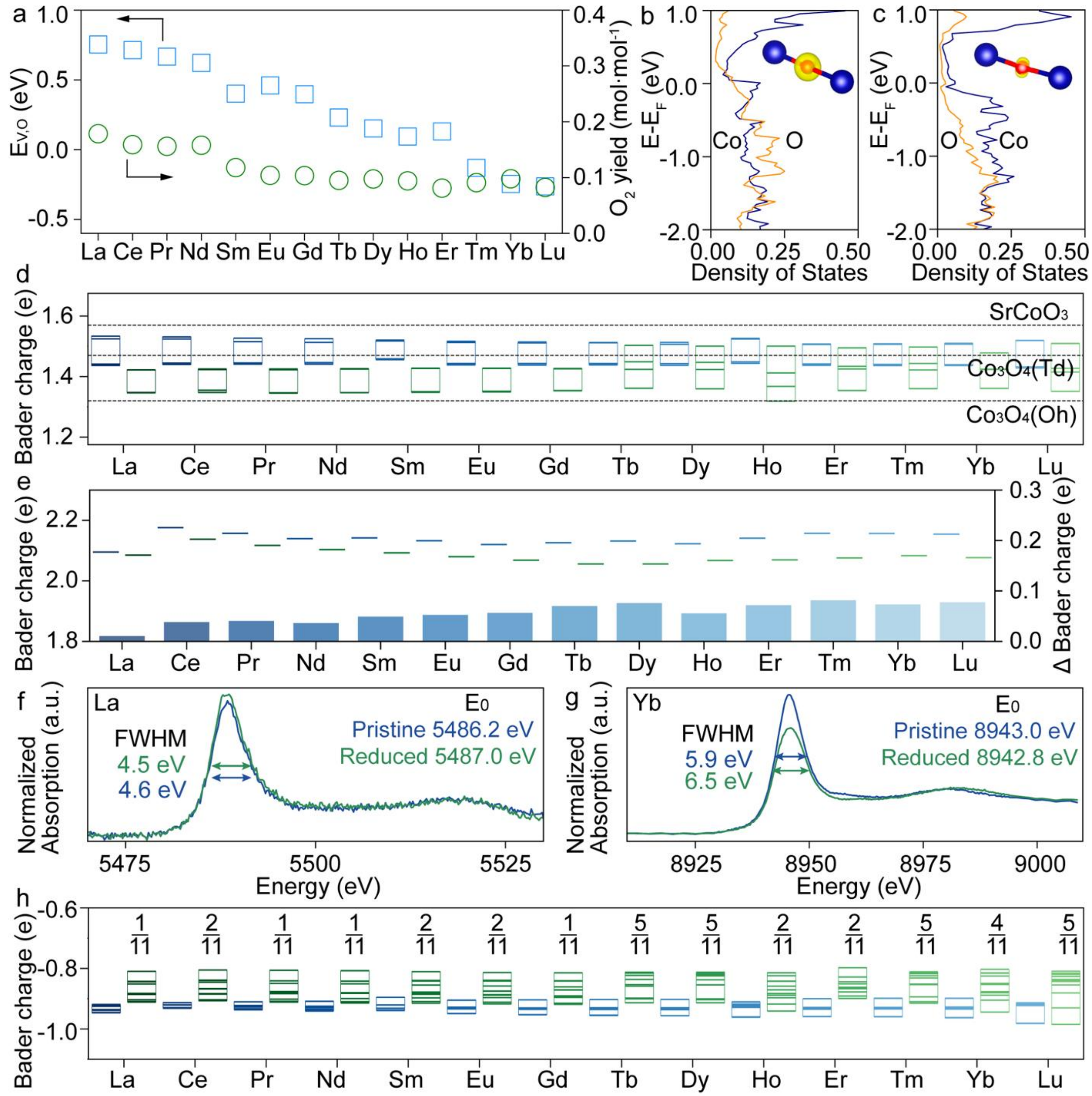

**Figure 3. Competing cationic and anionic charge-compensation pathways govern oxygen capacity. a**, Oxygen vacancy formation energy ($E_{V,O}$) and the cumulative $O_2$ yields of $Ln_{0.25}Sr_{0.75}CoO_3$ perovskites. **b** and **c**, The pDOS of Co 3*d* and O 2*p* in pristine and reduced $La_{0.25}Sr_{0.75}CoO_{3-\delta}$ perovskites; the inserted images show the PARCHG of O atoms. **d and e**, The Bader charge of Co and Ln in pristine and reduced $Ln_{0.25}Sr_{0.75}CoO_{3-\delta}$ perovskites. **f** and **g**, La and Yb $L_3$-edge XANES spectra, respectively, of pristine and reduced $Ln_{0.25}Sr_{0.75}CoO_{3-\delta}$ perovskites; the absorption edge energies ($E_0$) and full width at half maximum (FWHM) values are inserted. **h**, The Bader charge of O in pristine and reduced $Ln_{0.25}Sr_{0.75}CoO_{3-\delta}$ perovskites; the fractions of partially oxidized $O^{(2-\alpha)-}$ ($0<\alpha<2$) anions are inserted in Figure 3h.

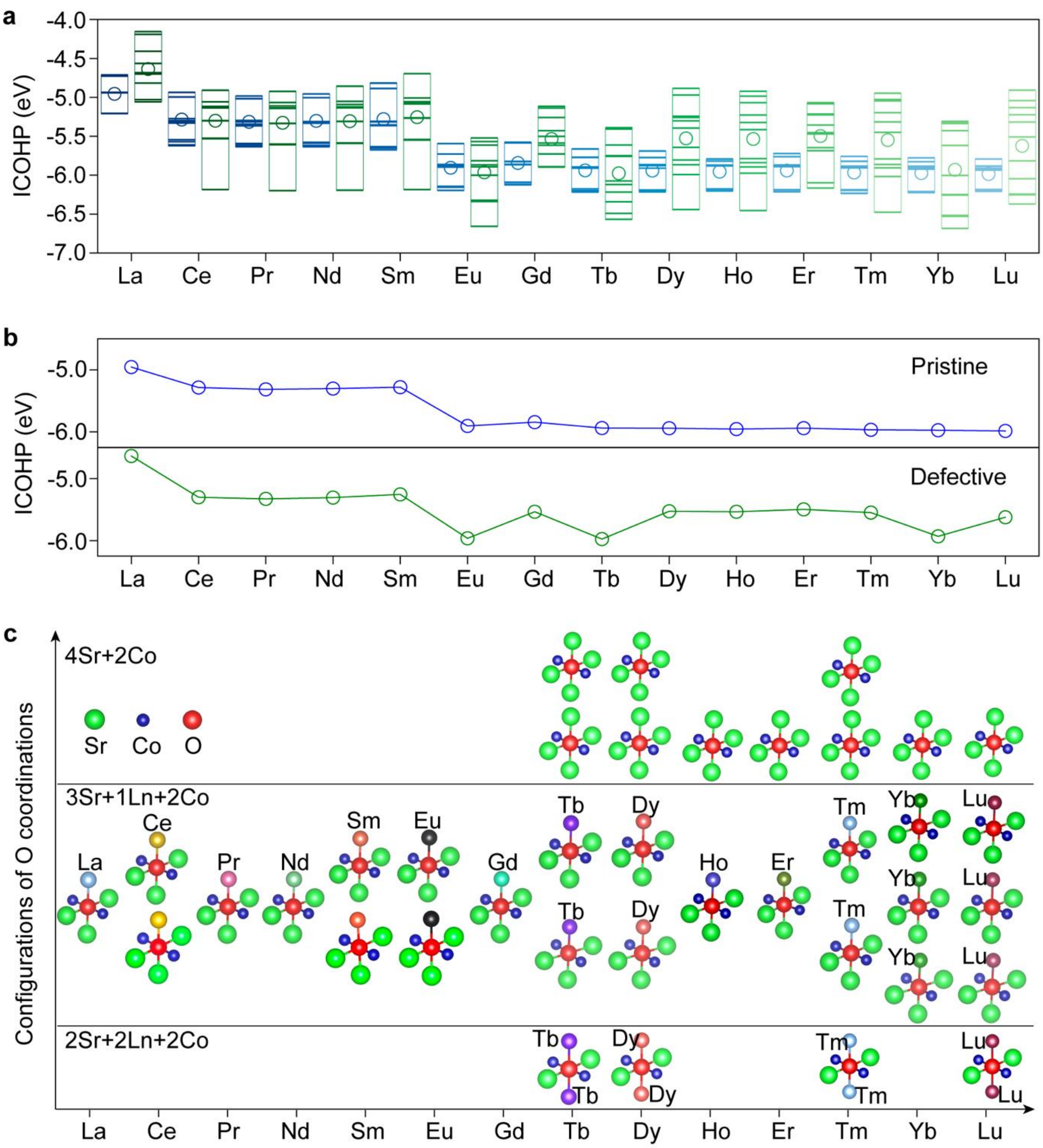

**Figure 4. Metal-oxygen bonding provides a descriptor for oxygen capacity. a**, The total M-O bonding interaction (ICOHP) distribution of each O in pristine and reduced $Ln_{0.25}Sr_{0.75}CoO_{3-\delta}$ perovskites. The circle markers inserted indicate the average values. **b**, The average values of total ICOHP of O in pristine and reduced $Ln_{0.25}Sr_{0.75}CoO_{3-\delta}$ perovskites. **c**, The coordination configurations of $O^{(2-\alpha)-}$ ($0<\alpha<2$) in reduced $Ln_{0.25}Sr_{0.75}CoO_{3-\delta}$ perovskites.